\documentclass[sigconf,natbib=false]{acmart}

\usepackage{balance}

\AtBeginDocument{%
  }

\RequirePackage[
  datamodel=acmdatamodel,
  style=acmnumeric,
  ]{biblatex}

\usepackage[most]{tcolorbox}
\tcbuselibrary{listingsutf8}
\usepackage{listings}
\usepackage{float}

\usepackage{multirow}
\usepackage{amsmath}

\begin{document}

\title{ReGeS: Reciprocal Retrieval–Generation Synergy for Conversational Recommender Systems}

\author{Dayu Yang}
\affiliation{%
    \institution{University of Delaware}
    \city{Newark}
    \state{DE}
    \country{USA}}
\email{dayu@udel.edu}

\author{Hui Fang}

\affiliation{%
  \institution{ University of Delaware}
  \city{Newark}
  \state{DE}
  \country{USA}}
\email{hfang@udel.edu}

\renewcommand{\shortauthors}{Dayu Yang, \& Hui Fang}

\begin{abstract}

Connecting conversation with external domain knowledge is vital for conversational recommender systems (CRS) to correctly understand user preferences. However, existing solutions either require domain-specific engineering, which limits flexibility, or rely solely on large language models, which increases the risk of hallucination. While Retrieval-Augmented Generation (RAG) holds promise, its naive use in CRS is hindered by noisy dialogues that weaken retrieval and by overlooked nuances among similar items. We propose \textbf{ReGeS}, a reciprocal \textbf{Re}trieval–\textbf{Ge}neration \textbf{S}ynergy framework that unifies generation-augmented retrieval to distill informative user intent from conversations and retrieval-augmented generation to differentiate subtle item features. This synergy obviates the need for extra annotations, reduces hallucinations, and simplifies continuous updates. Experiments on multiple CRS benchmarks show that ReGeS achieves state-of-the-art performance in recommendation accuracy, demonstrating the effectiveness of reciprocal synergy for knowledge-intensive CRS tasks. Our code is publicly available at the link: \url{https://github.com/dayuyang1999/ReGeS}

\end{abstract}

\ccsdesc[500]{Information systems~Users and interactive retrieval}
\ccsdesc[500]{Human-centered computing~HCI design and evaluation methods}

\keywords{Recommender Systems, Conversational Systems.}


\maketitle

\section{Introduction}

 Traditional Recommender Systems (RS) rely on passive signals like clicks or ratings to infer preferences, and therefore usually struggle when user preferences evolve~\cite{jannach2021survey}. Conversational Recommender Systems (CRS) address this by leveraging multi-turn dialogues to iteratively uncover user preferences and deliver tailored recommendations~\cite{zhou2020improving}. However, CRS must decode subtle user intents from casual speech, a task that demands extensive domain knowledge~\cite{gao2021advances}. To incorporate such knowledge, two main approaches have emerged. Representation-based methods inject domain-specific features into the pipeline~\cite{zhou2020improving, wang2022towards} but require intricate engineering and lack cross-domain flexibility. In contrast, Large Language Models (LLMs) serve as universal knowledge sources~\cite{he2023large, zhang2023user, feng2023large}; however, they must be continuously updated to accommodate new items and are prone to hallucinations, limiting their reliability in fast-evolving domains~\cite{ye2023cognitive, huang2024survey}.

\begin{figure}
    \centering
    \includegraphics[width=1\linewidth]{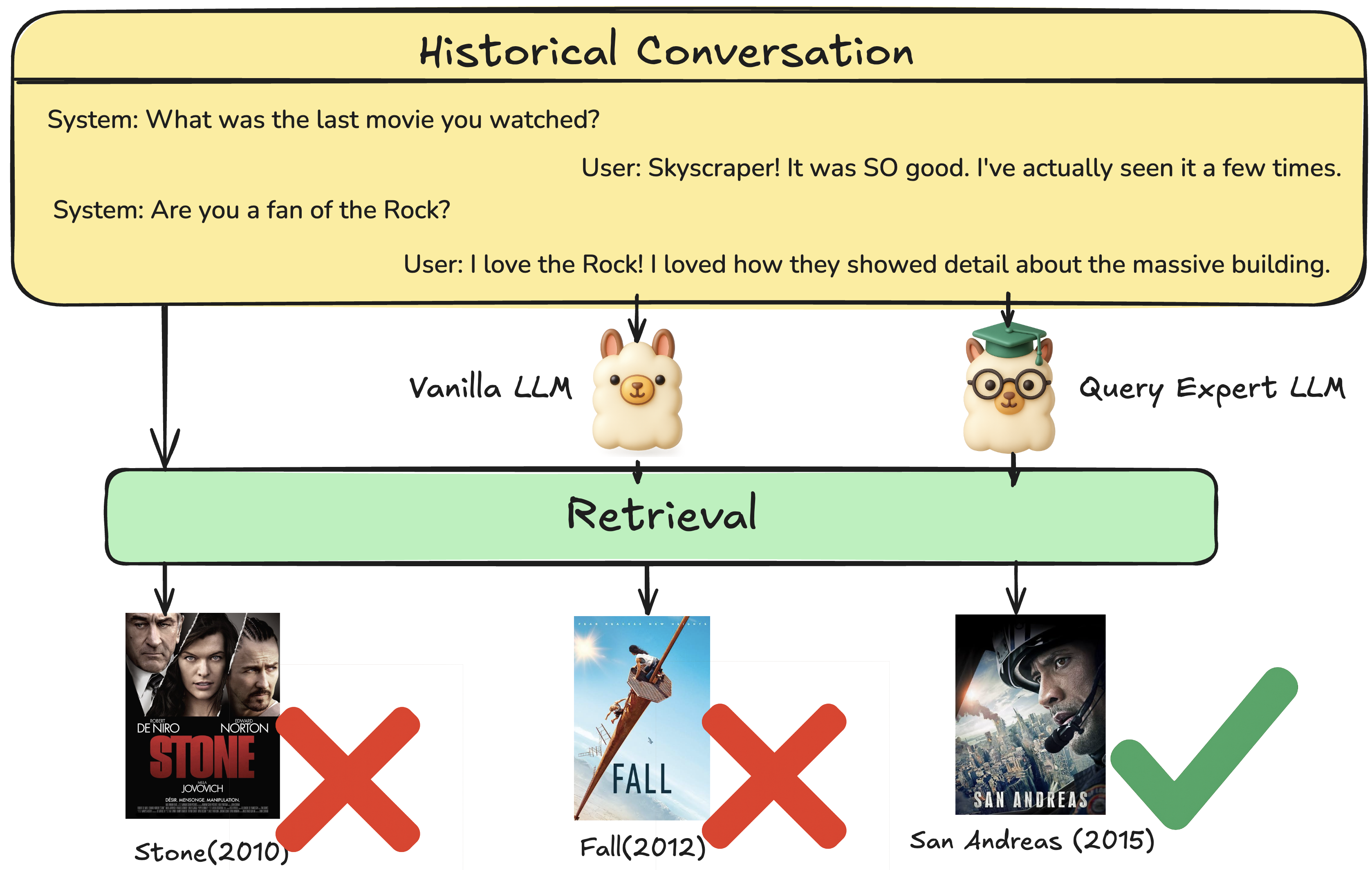}

    \caption{Given historical conversation in CRS, both directly using historical conversation and prompting vanilla LLMs fall short in retrieval.}
    
    \label{fig:query_reform_motivation}

\end{figure}
RAG has proven highly effective in knowledge-intensive NLP tasks by grounding outputs in external knowledge sources~\cite{lewis2020retrieval, xu2024retrieval, wu2024retrieval}. Given its strong capacity to inject factual information into conversational exchanges, one might expect RAG to be extensively studied in conversational recommendation systems (CRS). Surprisingly, it remains underexplored in this domain. Our initial attempts at applying RAG to CRS reveal two major hurdles (See experiment in \S~\ref{sec:g-e-r}): 1) \emph{noisy, lengthy conversation transcripts} impair retrieval effectiveness, and it cannot be easily solved by simply prompting LLMs as illustrated in Figure~\ref{fig:query_reform_motivation}. 2) \emph{suboptimal item generation} when confronted with multiple similar candidates, as LLMs struggle to differentiate subtle features.

Addressing the challenges of nosiy input and fine-grained item differentiation in isolation would be expensive, typically requiring extensive manual labeling~\cite{yang2023zero}. 

Specifically, training a model to denoise raw conversation transcripts is usually costly because it demands additional, annotated labels~\cite{hirsch2020query, chen2021towards}. In contrast, when coupled with a generative component, the LLM can leverage the ground-truth item to identify and extract the most salient concepts connecting the conversation to the desired recommendation, thereby generating high-quality supervision without human annotations. 


Furthermore, LLMs often fail to distinguish subtle differences among similar retrieved items~\cite{wang-etal-2024-large, gao2025llm4rerank}. Addressing this via hard negative training requires an effective retrieval stage capable of identifying these crucial near-miss examples. Initial poor retrieval performance thus obstructs effective generator training. This motivates a synergistic approach.


We propose \textbf{ReGeS}, a domain-agnostic framework that synergizes retrieval and generation for CRS. ReGeS introduces: 1) \textbf{generation-augmented retrieval}, where a trained LLM performs as the query expert to produce concise, preference-focused queries for a retrieve by leveraging the generated pseudo guiding signal. 2) \textbf{retrieval-augmented item generation}, where retrieved candidates train the LLM to discern subtle item distinctions. Our contributions are: 1) the first RAG-based CRS framework integrating grounded knowledge retrieval; 2) a reciprocal pipeline that synergistically integrates retrieval and generation to overcome the challenges of noisy input and item ambiguity without costly annotations; and 3) extensive experiments that ReGeS significantly outperforms state-of-the-art methods, mitigating hallucination risks and avoiding laborious domain-specific engineering.

\begin{figure*}
    
    \centering
    \includegraphics[width=1\linewidth]{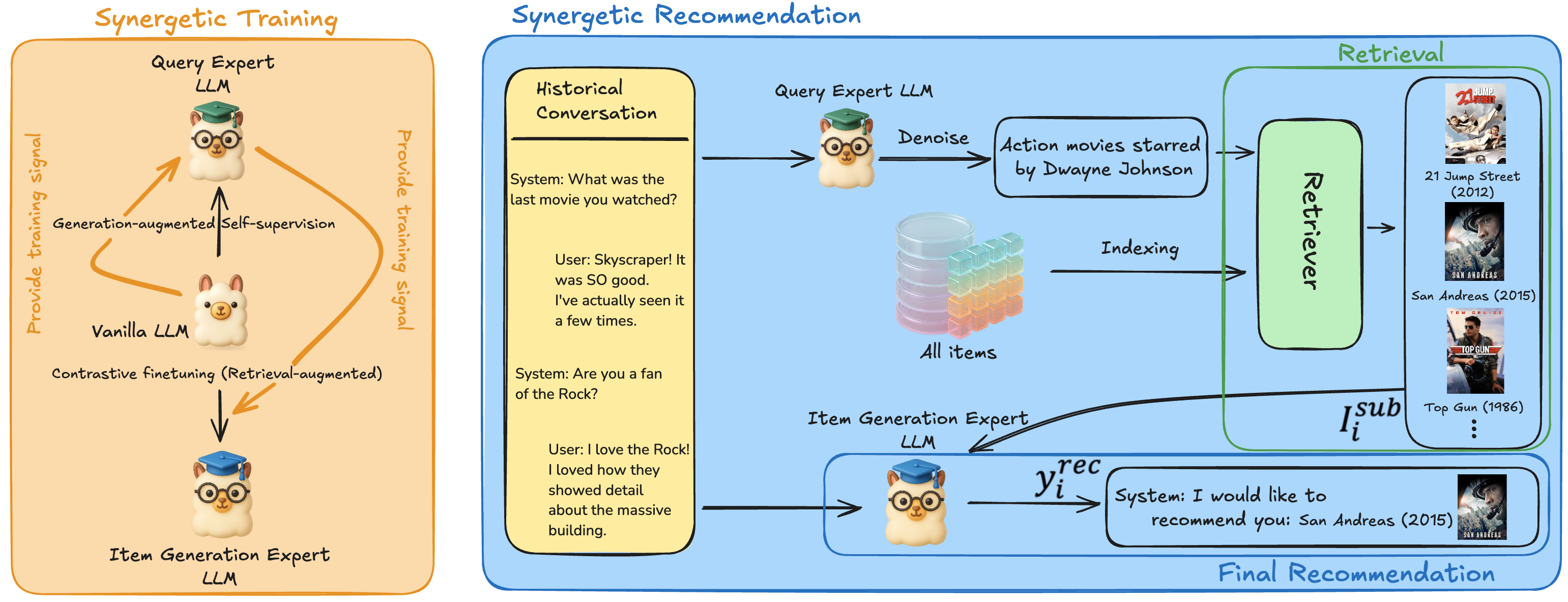}
    \caption{Overview of the ReGeS framework, illustrating the reciprocal synergy. Left (Training): Generation guides Query Expert $LLM_{QR}$ training, while Retrieval provides training signals for Item Generation Expert $LLM_G$ training. Right (Inference): The trained components collaborate to produce the recommendation.}
    \label{fig:workflow}

\end{figure*}

\section{Methodology}
\label{sec:methodology}

In this section, we introduce the methodological detail of ReGeS. \S~\ref{sec:overview} provides an overview of how these two components collaborate, followed by a formal problem definition in \S~\ref{sec:problem}. We then present two core modules: \textit{Generation-augmented Retrieval} (\S~\ref{sec:gen_augmented_retrieval}) and \textit{Retrieval-augmented Generation} (\S~\ref{sec:ret_augmented_generation}). This bidirectional flow—where generation refines retrieval and retrieval sharpens generation—ensures final recommendations are both contextually grounded and factually accurate.

\subsection{Overview}
\label{sec:overview}

Figure~\ref{fig:workflow} illustrates the reciprocal interplay of retrieval and generation in ReGeS. At the retrieval stage, a \textit{query expert LLM} first condenses a user--system dialogue history $\mathcal{H}_i$ into a concise, preference-aware query $\hat{q}_i$. By leveraging grounded user-system interaction to generate pseudo supervision signal as self-supervision, query expert LLM learns which aspects of $\mathcal{H}_i$ matter for retrieval, alleviating the challenge of noisy and lengthy conversations. 
During generation, the predicted informative query $\hat{q}_i$ is used by a retriever $R$ to select candidate items $\mathcal{I}^{\text{sub}}_i$. Given $\mathcal{H}_i$ and these retrieval results (including closely-related ``hard negatives''), a contrastively trained generator $G$ pinpoints the most suitable recommendation $y^{\text{rec}}_i \in \mathcal{I}^{\text{sub}}_i$.

\subsection{Problem Statement}
\label{sec:problem}

The problem of CRS can be expressed as learning a mapping between the $i$-th user-system interaction history $\mathcal{H}_i$ and the ground-truth item $y_i$, given the candidate set $\mathcal{I}$:
\begin{equation}\notag
    y^{\text{rec}}_i = CRS(\mathcal{H}_i; \mathcal{I}),
\end{equation}
where $y_i^{\text{rec}}$ is the recommended (predicted) item that should ideally be the same as $y_i$. $\mathcal{I} = \{i_1, i_2, \dots, i_M\}$ is the set of all possible items that are agnostic to users. Typically, a Retrieval-Augmented Generation paradigm that decomposes the process into two steps:
\begin{equation}\notag
\mathcal{I}^{\text{sub}}_i = R(\mathcal{H}_i, \mathcal{I}),
\label{eq:retrieval}
\end{equation}
\begin{equation}\notag
y^{\text{rec}}_i = G(\mathcal{H}_i, \mathcal{I}^{\text{sub}}_i),
\label{eq:generation}
\end{equation}
where $R$ retrieves a relevant candidate set, and $G$ generates the final recommendation. Despite this formulation’s appeal, two obstacles hinder CRS effectiveness: (1)~noisy conversational input complicates retrieval, and (2)~hard-to-distinguish item candidates challenge generation. In what follows, we describe how ReGeS solves both issues by leveraging a circular flow of information between retrieval and generation.

\subsection{Generation-Augmented Retrieval}
\label{sec:gen_augmented_retrieval}

\begin{figure}
    \centering
    \includegraphics[width=\linewidth]{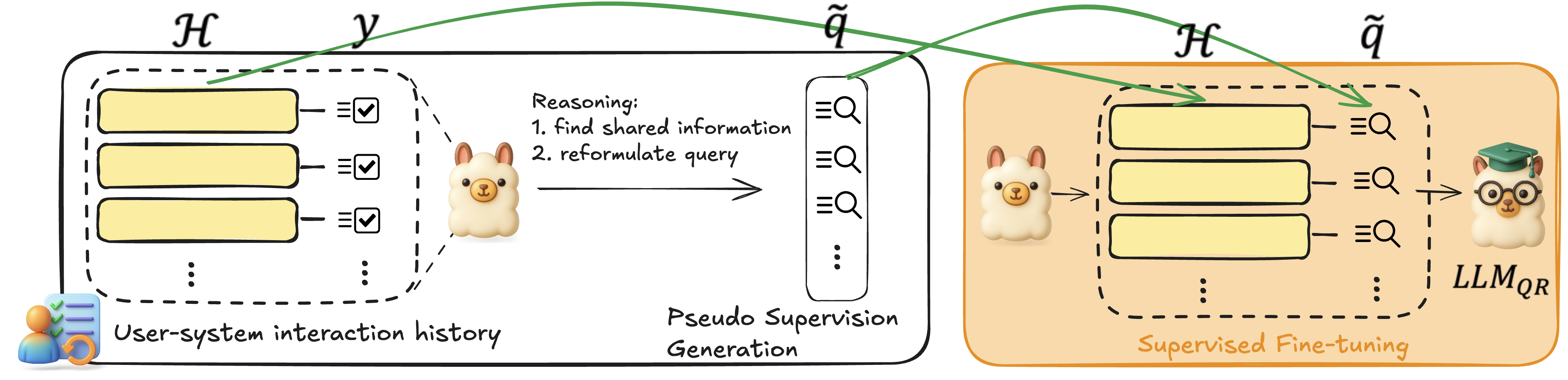}
    \caption{Process of building generation-augmented query expert $LLM_{QR}$ via generative self-supervision.}
    \label{fig:GER}
\end{figure}

To improve retrieval performance under noisy input, we introduce a \emph{generation-augmented query expert LLM}, as shown in Figure~\ref{fig:GER}, that learns to produce a concise, information-rich query $\hat{q}_i$ from the conversation history $\mathcal{H}_i$. This module is built in a self-supervised manner, alleviating the need for costly human annotations.

\noindent

\noindent\textbf{Generative Self-Supervision via Ground-Truth Guidance.} Given a training instance with conversation  and its ground-truth recommended item $(\mathcal{H}_i, y_i)$, we first prompt a LLM to produce an informative query $\tilde{q}_i$:
\[
\tilde{q}_i = \mathrm{LLM}\bigl(\mathcal{H}_i, y_i\bigr).
\]
Because $y_i$ is provided, the LLM focuses on the essential features in $\mathcal{H}_i$ that justify why $y_i$ is recommended, thus generating $\tilde{q}_i$ as a pseudo-labeled query without manual annotations. 

Next, we fine-tune a second LLM instance, denoted $\mathrm{LLM}_{\mathrm{QR}}$, to reproduce $\tilde{q}_i$ \emph{without} access to $y_i$. Formally, we minimize the cross-entropy loss:
\[
\mathcal{L}_{\mathrm{QR}} = 
\sum_{i=1}^{N} \mathcal{L}_{\mathrm{CE}}\Bigl(
\mathrm{LLM}_{\mathrm{QR}}(\mathcal{H}_i), 
\tilde{q}_i \Bigr),
\]
over the training set. At inference time, $LLM_{QR}$ transforms $\mathcal{H}_i$ into a compact query $\hat{q}_i = \mathrm{LLM}_{\mathrm{QR}}(\mathcal{H}_i)$ capturing the most salient user preferences.

\noindent\textbf{Refined Retrieval.} The refined query $\hat{q}_i$ serves as input to the retriever $R$, which selects a subset of relevant items:
\[
\mathcal{I}^{\text{sub}}_i = R\bigl(\hat{q}_i, \mathcal{I}\bigr).
\]
By disentangling user intents from extraneous chatter, $LLM_{QR}$ substantially boosts retrieval accuracy. Moreover, $LLM_{QR}$ itself is \textit{generation-augmented}---it relies on knowledge gleaned from the final item $y_i$, harnessing generation to improve retrieval.

\subsection{Retrieval-Augmented Generation}
\label{sec:ret_augmented_generation}

With $\mathcal{I}^{\text{sub}}_i$ now distilled to the most relevant candidates, the next step is to generate the final recommendation $y^{\text{rec}}_i$. A pure text-generation approach could overlook subtle differences among items, leading to hallucinated or incorrect picks, especially when multiple items in $\mathcal{I}^{\text{sub}}_i$ are nearly interchangeable. ReGeS avoids this pitfall by feeding the retrieved candidates back into the generation process in a contrastive manner, ensuring the generator can pinpoint the precise match. 

\noindent\textbf{Retrieve Challenging Negatives with $LLM_{QR}$.} During training, each instance consists of $\mathcal{H}_i$, its ground-truth item $y_i$, and a set of top-$k$ retrieved items $\{c_{i_1},\ldots,c_{i_k}\}$ utilizing $LLM_{QR}$ that is previously built, where none of the elements is $y_i$. Because of the improved retrieval performance brought by $LLM_{QR}$, these ``hard negatives'' will share strong relevance signals with $\mathcal{H}_i$, they elevate challenges the Item Generation Expert $LLM_{Q}$ to distinguish subtle nuances. Let $\mathcal{S}_i$ be the item set we fed into LLM during training, i.e.,
\[
\mathcal{S}_i = \{\, y_i,\, c_{i_1},\, \dots,\, c_{i_k} \},
\]
$|\mathcal{S}_i|= k+1$ for all training instances.
The model sees $\mathcal{H}_i, \mathcal{S}_i$, and a prompt specifying that only $y_i$ is correct.

\noindent\textbf{Contrastive Fine-Tuning.} We then fine-tune an LLM-based generator $LLM_{G}$ to maximize the likelihood of selecting $y_i$. Formally, we encourage
\[
P\bigl(\text{text}(y_i)\mid \mathcal{H}_i, \mathcal{S}_i\bigr) \;>\;
P\bigl(\text{text}(c_{i_k})\mid \mathcal{H}_i, \mathcal{S}_i\bigr),
\]
for all $c_{i_k}$. 
We implement this using a standard cross-entropy loss, maximizing the probability assigned to the ground-truth item $y_i$ within the context of the candidate set $S_i$. By confronting LLM with multiple lookalike items, it acquires a fine-grained capability to identify the correct candidate.

\noindent\textbf{Final Recommendation.} At inference time, $LLM_{G}$ receives both $\mathcal{H}_i$ and the retrieved subset $\mathcal{I}^{\text{sub}}_i$. Because the generator has learned to discriminate items using hard negatives, it selects the true best item $y^{\text{rec}}_i\in \mathcal{I}^{\text{sub}}_i$ with higher fidelity. Consequently, retrieval is \textit{generation-augmented}, and generation is \textit{retrieval-augmented}, creating a bidirectional synergy that overcomes the pitfalls of either component in isolation.

\begin{figure}
    \centering
    \includegraphics[width=1.0\linewidth]{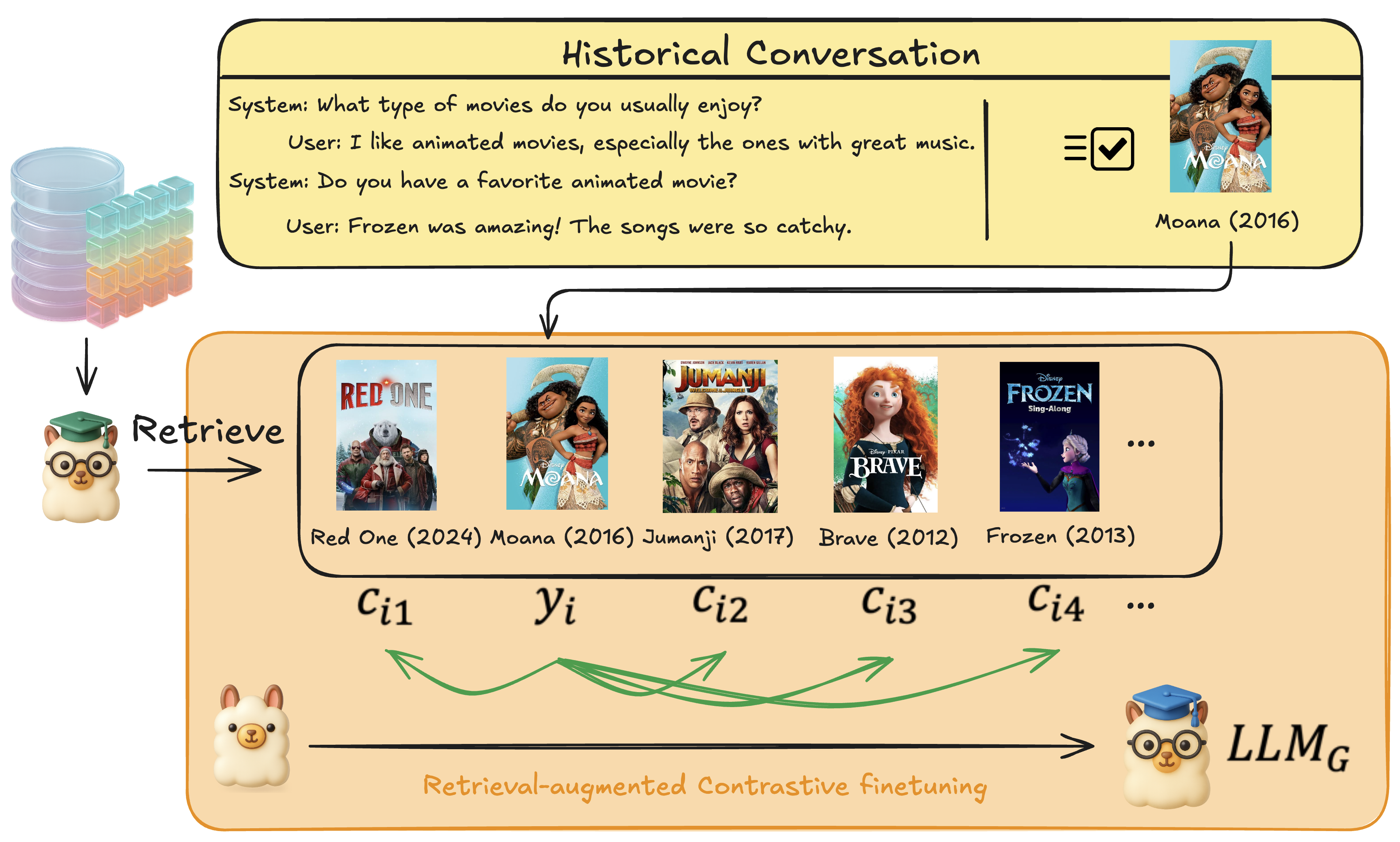}
    \caption{Retrieval-augmented Contrastive fine-tuning for building Item Generation Expert $LLM_{G}$.}
    \label{fig:REG}
\end{figure}

\section{Experiments}
\label{sec:exp}

In this section, we first introduce the benchmark datasets, the baselines and experimental settings in \S\ref{sec:datasets}, \S\ref{sec:baselines} and \S\ref{sec:exp_settings} respectively. We then present the overall results in \S\ref{sec:overall}, show ablation studies in \S\ref{sec:ablation}, and conduct additional analyses, including chain-of-thought generation in \S\ref{sec:cot}.

\subsection{Datasets}
\label{sec:datasets}

Consistent with most previous CRS studies~\cite{li2018towards, chen2019towards,zhou2020improving, wang2022towards, feng2023large}, we conduct experiments on two widely-used CRS datasets, \textbf{ReDial} \cite{li2018towards} and \textbf{INSPIRED}~\cite{hayati2020inspired}. 
ReDial contains 10,006 training and 1,342 test dialogs, focusing on movie recommendations. While relatively large, each conversation tends to be shorter, offering limited explicit preference cues. 
INSPIRED contains 801 training and 198 validation/test dialogs with richer preference interactions. Both datasets provide ground-truth recommended movies for each conversation, which we treat as the target item(s).

Following prior work \cite{zhou2021crslab,wang2022towards}, we apply the same preprocessing pipelines and focus on the \emph{recommendation} turn in each dialogue.

\subsection{Baselines}
\label{sec:baselines}

We compare our proposed \textbf{ReGeS} with three categories of conversational recommendation methods:

    \textbf{Traditional Representation-Based Methods.} 
    \textit{KBRD} \cite{chen2019towards} integrates DBpedia knowledge for semantic enhancement, while \textit{KGSF} \cite{zhou2020improving} leverages both DBpedia and ConceptNet for enriched reasoning. \textit{UniCRS} \cite{wang2022towards} encodes item meta-information in prompt embeddings for unified modeling of user--item interactions.
    
    \textbf{LLM-Based Methods (LLM-CRS).}
    \textit{GPT-3.5-turbo} and \textit{GPT-4} \cite{he2023large,hou2024large} are closed-source LLMs, while \textit{Vicuna-13B} \cite{chiang2023vicuna} is open-source. Each generates recommendations by analyzing the conversation history without explicit retrieval-based updates.

    \textbf{RAG Methods directly borrow from Recommender System (RAG-RS).}
    \textit{RARS}~\cite{di2023retrieval} Assuming high-quality user query is provided for Retreiver and prompt vanilla LLM with retrieved information to generate recommendations. \textit{RAMO}~\cite{rao2024ramo} In addition to user query, relevant document for mentioned items are also provided for the retrieval stage.

\subsection{Implementation Details}
\label{sec:exp_settings}

\paragraph{Models and Retrievers}
We instantiate ReGeS using three sizes of Large Language Models (LLMs): Gemma-2B~\cite{team2024gemma}, LLaMA3.1-8B~\cite{metaai2024}, and Gemma-27B~\cite{team2024gemma}. For the retrieval component, we evaluate three dense retrievers with varying capacities: DPR~\cite{karpukhin2020dense} (110M parameters, 768-dim embeddings), BGE-large-1.5-en~\cite{bge_embedding} (BGE, 340M parameters, 1024-dim embeddings), and OpenAI's text-embedding-3-large~\cite{openai2023} (3074-dim embeddings). The retriever returns the top-50 candidates ($k=50$) for the generation stage; an analysis of the impact of $k$ is provided in \S \ref{appendix:num_candidates}.

\paragraph{ReGeS Training Protocol}
We detail the fine-tuning procedures for the two core LLM components of ReGeS:

\textit{1) Generation-Augmented query expert ($\mathrm{LLM}_{\mathrm{QR}}$):} the LLM is fine-tuned using the pseudo-queries generated via self-supervision (\S\ref{sec:gen_augmented_retrieval}). For Gemma-2B and Llama3.1-8B, we train for one epoch with a batch size of 2. For the larger Gemma-27B model, we use a batch size of 1 for one epoch. We employ the AdamW optimizer with learning rates of 1e-4 for the 2B/8B models and 2e-5 for the 27B model. A linear learning rate warmup schedule is used over the first 10\% of training steps.

\textit{2) Retrieval-Augmented Item Generation Expert ($\mathrm{LLM}_{G}$):} The item generation LLM is fine-tuned using a contrastive objective with hard negatives supplied by the retriever (\S\ref{sec:ret_augmented_generation}). One epoch with a batch size of 2 (2B/8B) or 1 (27B). The AdamW optimizer is used with learning rates of 1e-4 (2B/8B) and 5e-5 (27B), again with a 10\% warmup phase.

\paragraph{Technical Specifications}
To manage computational resources effectively, we set the maximum input sequence length to 4096 tokens, truncating longer sequences. We utilize Low-Rank Adaptation (LoRA)~\cite{hu2021lora} with rank $r=8$ and scaling factor $\alpha=32$, applied only to the query and value projection matrices within the self-attention mechanism of the LLMs. During inference for item generation, we set the decoding temperature to 0.1 to promote deterministic and reproducible outputs. All experiments were conducted on a single NVIDIA H100 80GB GPU.

\paragraph{Prompt Templates}
The prompts used for generating pseudo-supervision for $\mathrm{LLM}_{\mathrm{QR}}$ and for the contrastive fine-tuning of $\mathrm{LLM}_{G}$ are crucial for reproducibility. The specific templates are provided below. Placeholders like \texttt{<conversation\_history>} and \texttt{<item\_name>} are dynamically filled during training and inference.

\begin{tcolorbox}[title=Prompt for \texttt{LLM\textsubscript{QR}} Self-Supervision, 
                  colback=gray!5, colframe=black!20, fonttitle=\bfseries]
\begin{lstlisting}
Find the common concepts between the conversation history that the seeker reveals and the ground truth movie, particularly paying attention to actor, producer, genre, topics, etc. Then, formulate a good natural-language search query for matching introduction of movies the seeker likes. Only output the query, do not include any explanation.

Conversation history: <conversation_history>

Ground truth items: <ground_truth_item>
\end{lstlisting}
\end{tcolorbox}


\begin{tcolorbox}[title=Prompt for \texttt{LLM\textsubscript{G}} Contrastive Fine-Tuning/Inference, 
                  colback=gray!5, colframe=black!20, fonttitle=\bfseries]
\begin{lstlisting}
A list of candidate movies and their abstracts are provided. According to the conversation history between seeker and recommender, select the top movie recommendation for the seeker from the candidate list. Only output one movie name. Do not generate explanation or anything else.

A list of candidate abstracts: <item_1> <item_2>, ..., <item_n>

The corresponding conversation history: <conversation_history>
\end{lstlisting}
\end{tcolorbox}

\paragraph{Baseline Implementations}
For baseline comparisons, we implemented UniCRS~\cite{wang2022towards}, KGSF~\cite{zhou2020improving}, and KBRD~\cite{chen2019towards} using their publicly available code and recommended hyperparameters. For LLM-based baselines (GPT-3.5-turbo, GPT-4, Vicuna-13B), we followed the experimental setups and prompts described in prior work~\cite{he2023large, hou2024large, chiang2023vicuna}. As code for RAMO~\cite{rao2024ramo} and RARS~\cite{di2023retrieval} was not available, we re-implemented their core methodologies using DPR and Llama3.1-8B for a fair comparison within our experimental setup.

\paragraph{Evaluation Metric}
Our primary evaluation metric is the Recommendation Success Rate (Rec Success Rate in Table~\ref{tab:overall_results}), equivalent to Recall@1 or Hit@1, focusing on the accuracy of the final item recommendation. We deliberately omit traditional natural language generation metrics (e.g., BLEU, ROUGE, DIST). While historically relevant for assessing the fluency challenges of earlier CRS models \cite{chen2019towards, zhou2020improving, wang2022towards}, these metrics offer diminishing returns for evaluating modern LLMs, which consistently produce high-quality, coherent text \cite{mendoncca2024benchmarking, gao2024llm}. Such n-gram based metrics are often saturated and fail to capture the semantic correctness or relevance of the recommendation itself \cite{wang2023rethinking, yang2024behavior}. Therefore, focusing on Rec Success Rate provides a more direct and meaningful assessment of the core recommendation task in contemporary CRS.

\subsection{Overall Performance}
\label{sec:overall}

\begin{table}[ht] 
\centering 
\resizebox{.5\textwidth}{!}{%
\begin{tabular}{ccccc}
\hline
\multicolumn{3}{l}{\multirow{1}{*}{\textbf{Method}}}                                      & \textbf{ReDial}      & \textbf{INSPIRED}    \\ \hline \hline
\multicolumn{1}{l}{\textbf{Representation-CRS}}     & \multicolumn{2}{l}{}                                                                      & \multicolumn{2}{c}{\textbf{Rec Success Rate}}   \\ \hline
\multicolumn{3}{c|}{KBRD}                                                                 & 0.027                & 0.036                \\
\multicolumn{3}{c|}{KGSF}                                                                 & 0.030                & 0.051                \\
\multicolumn{3}{c|}{UniCRS}                              & 0.049                & 0.041\textsuperscript{\ref{fn:unicrs}}                \\ \hline
\multicolumn{1}{l}{\textbf{LLM-CRS}}     & \multicolumn{1}{l}{} & \multicolumn{1}{l}{}     & \multicolumn{1}{l}{} & \multicolumn{1}{l}{} \\ \hline
\multicolumn{3}{c|}{GPT-3.5}                                                              & 0.041                & 0.047                \\
\multicolumn{3}{c|}{GPT-4}                                                                & 0.043                & 0.062                \\
\multicolumn{3}{c|}{Vicuna-13B}                                                          & 0.031                & 0.067                \\ \hline
\multicolumn{1}{l}{\textbf{RAG-RS}}     & \multicolumn{1}{l}{} & \multicolumn{1}{l}{}     & \multicolumn{1}{l}{} & \multicolumn{1}{l}{} \\ \hline
\multicolumn{3}{c|}{RAMO}                                                                 & 0.017                & 0.039                \\
\multicolumn{3}{c|}{RARS}                                                                 & 0.041                & 0.059                \\ \hline
\multicolumn{1}{l}{\textbf{ReGeS}} & \multicolumn{1}{l}{} & \multicolumn{1}{l}{}     & \multicolumn{1}{l}{} & \multicolumn{1}{l}{} \\ \hline
Computational Budget                    & Retriever            & Generation               & \multicolumn{1}{l}{} & \multicolumn{1}{l}{} \\ \hline
Small                                   & DPR                  & \multicolumn{1}{c|}{2B}  & 0.054\textsuperscript{\textdagger} & 0.108\textsuperscript{\textdagger} \\ 
Middle                                  & BGE                  & \multicolumn{1}{c|}{8B}  & 0.084                & 0.196                \\
Large                                   & OpenAI               & \multicolumn{1}{c|}{27B} & 0.083                & 0.137                \\ \hline
Best (CoT)                              & OpenAI               & \multicolumn{1}{c|}{8B}  & \textbf{0.094}       & \textbf{0.218}       \\ \hline
\end{tabular}%
}
\caption{Recommendation success rate on the two datasets. Bold indicates the best overall result. The dagger (\textdagger) marks the ReGeS (Small) setting; statistical significance tests (e.g., paired t-test, p < 0.01) confirm that even this weakest ReGeS configuration significantly outperforms the strongest baselines from each category (UniCRS, GPT-4, RARS) on both datasets.} 
\label{tab:overall_results}
\end{table}

\footnotetext[4]{\label{fn:unicrs}Value obtained through reproduction using authors' published code and recommended hyperparameters. Original reported value was 0.094, but multiple independent reproductions achieved 0.041.}

\begin{figure*}
    \centering
    \includegraphics[width=1.0\linewidth]{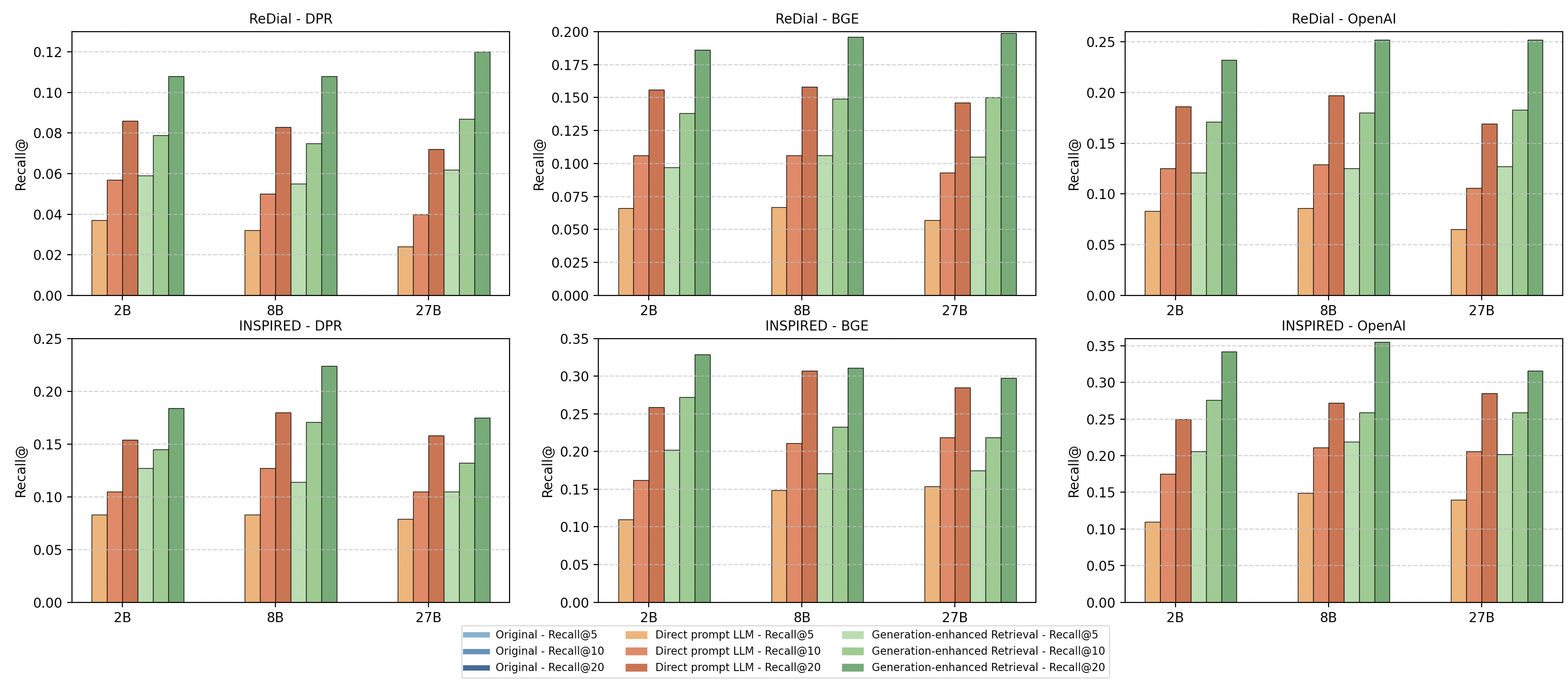}
    \caption{Retrieval performance comparison between our method (Green) and other baseline methods across different queries and retrievers on both datasets. Note that the scales of y-axis may be different.}
    \label{fig:full_ablation_exp1}
\end{figure*}

Table~\ref{tab:overall_results} presents the recommendation performance of our proposed method, ReGeS, compared to the baseline methods. We showcase three Retriever-LLM combinations in the table, each representing different levels of computational budget. Notably, performance does not always improve with larger retrievers and LLMs. This may be due to the fact that the Llama3.1 models are newer and more efficient than the older Gemma models. We also highlight the best overall performance of ReGeS, achieved when using Chain-of-Thought (CoT) item generation, the text-embedding-large-3 retriever, and Llama3.1-8B. By default, CoT is not used.

Compared to traditional graph-based methods like KBRD and KGSF, ReGeS directly retrieves and filters item knowledge from a broader pool, obviating the need for separate graph modules. ReGeS also outperforms purely LLM-based systems (GPT-3.5-turbo, GPT-4, and Vicuna-13B), highlighting the value of retrieval-augmentation in handling domain-specific updates and reducing hallucinations. Moreover, compared to existing \textbf{RAG-RS} methods (RAMO, RARS), our approach substantially boosts accuracy, underscoring the importance of \textit{reciprocal} synergy, as we will discuss in the next subsection.

\subsection{Effect of Reciprocal Synergy}
\label{sec:ablation}

We now examine how ReGeS's two core components: Generation-Augmented Retrieval (G-augmented R) and Retrieval-Augmented Generation (R-augmented G) address their respective challenges and merge into a synergetic framework.


\paragraph{Generation-Augmented Retrieval.}
\label{sec:g-e-r}


Figure~\ref{fig:full_ablation_exp1} illustrates how each query denoising method impacts retrieval performance on INSPIRED. We highlight three key observations:

    \textit{1) Query Quality Is Crucial.} When using the unprocessed \textit{Original} conversation as the query, all retrievers underperform markedly. For instance, with the DPR retriever, the Recall@5 is only 0.035, meaning just 3.5\% of the top five retrieved items include the ground-truth recommendation.
    
    \textit{2) Direct Prompt Alleviates Some Noise.} Prompting a vanilla LLM to produce a shorter query helps increase retrieval metrics compared to \textit{Original}. However, \textit{Direct Prompt} still falls short when crucial details are inadvertently omitted, emphasizing the importance of a careful query refinement.
    
    \textit{3) G-Augmented R Achieves Consistently Superior Results.} Our approach substantially outperforms the other two strategies. On average, Recall@20 improves by 47\% over \textit{Original} and 26\% over \textit{Direct Prompt}, while Recall@5 sees gains of 81\% and 60\%.  This improvement arises because the self-refinement process systematically identifies the key attributes linking the conversation to the ground-truth item, rather than preserving noise.

\paragraph{Retrieval-Augmented Generation.}
\label{sec:r-e-g}

\begin{figure}
    \centering
    \includegraphics[width=1\linewidth]{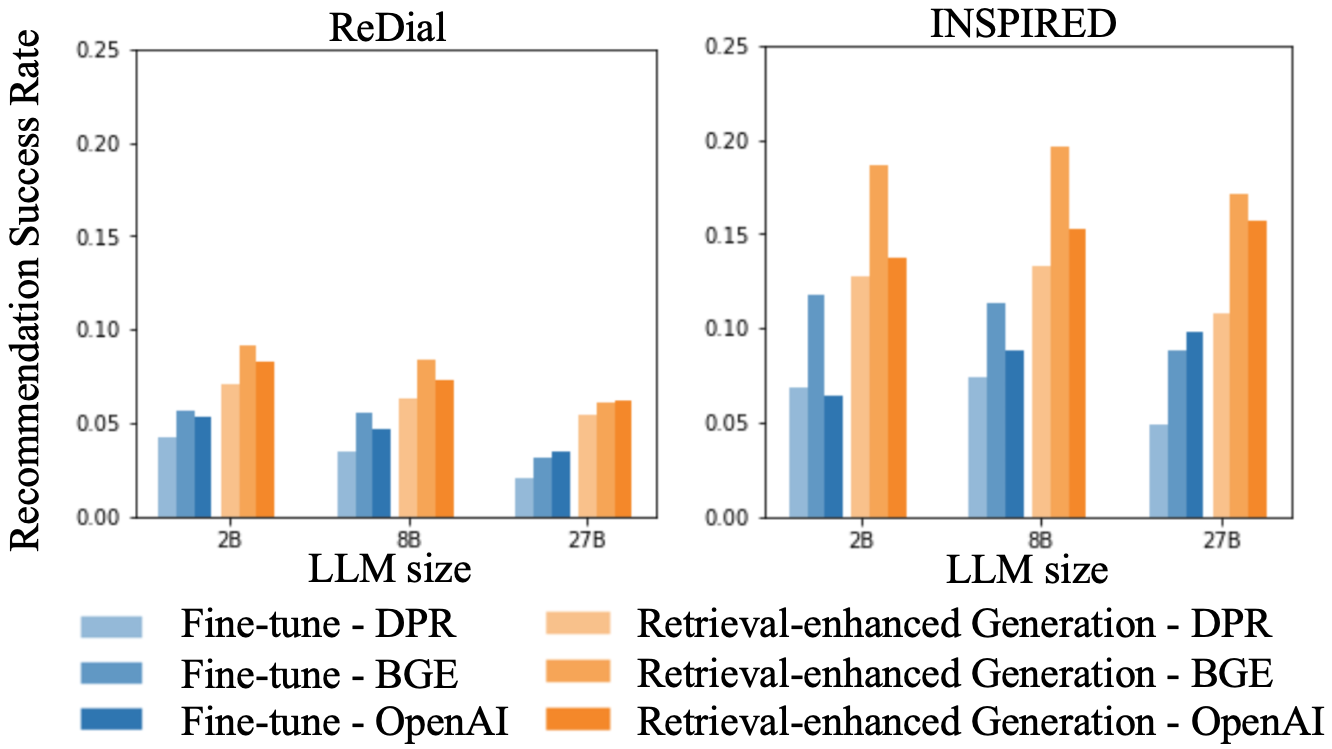}
    \caption{Item recommendation performance comparison between our method (Orange) and baseline methods (Blue) across various RAG settings.}
    \label{fig:ablation_exp2}
\end{figure}


Figure~\ref{fig:ablation_exp2} compares the item recommendation success rate of our proposed R-augmented G method with the baseline method "fine-tune", which fine-tunes the LLM without incorporating hard negatives obtained from the retriever into the training data. On average, incorporating hard negatives boosts recommendation success rate by \textbf{79\%} on ReDial and \textbf{85\%} on INSPIRED. This underlines the importance of bridging the distribution gap between training data (where random negatives might not resemble real-world distractors) and actual inference data (where retrieval returns plausible, similar candidates). R-augmented G mitigates hallucination and confusion, thereby reducing errors in the final recommendation step.

\subsection{Additional Analyses}
\label{sec:cot}

This section shows how two obstacles hinder RAG-CRS effectiveness and how G-augmented R and R-augmented G overcome them.

\subsubsection{Generation-augmented Retrieval: Improving Query Quality}

We hypothesize that G-augmented R will yield queries that are both semantically closer to user-preferred items and more aligned with human annotaions. To test this, we conduct two primary analyses: 1) measuring how queries from different methods compared in the latent space to ground-truth items, and 2) evaluating textual similarity against human-annotated queries.

\begin{figure}
    \centering
    \includegraphics[width=1\linewidth]{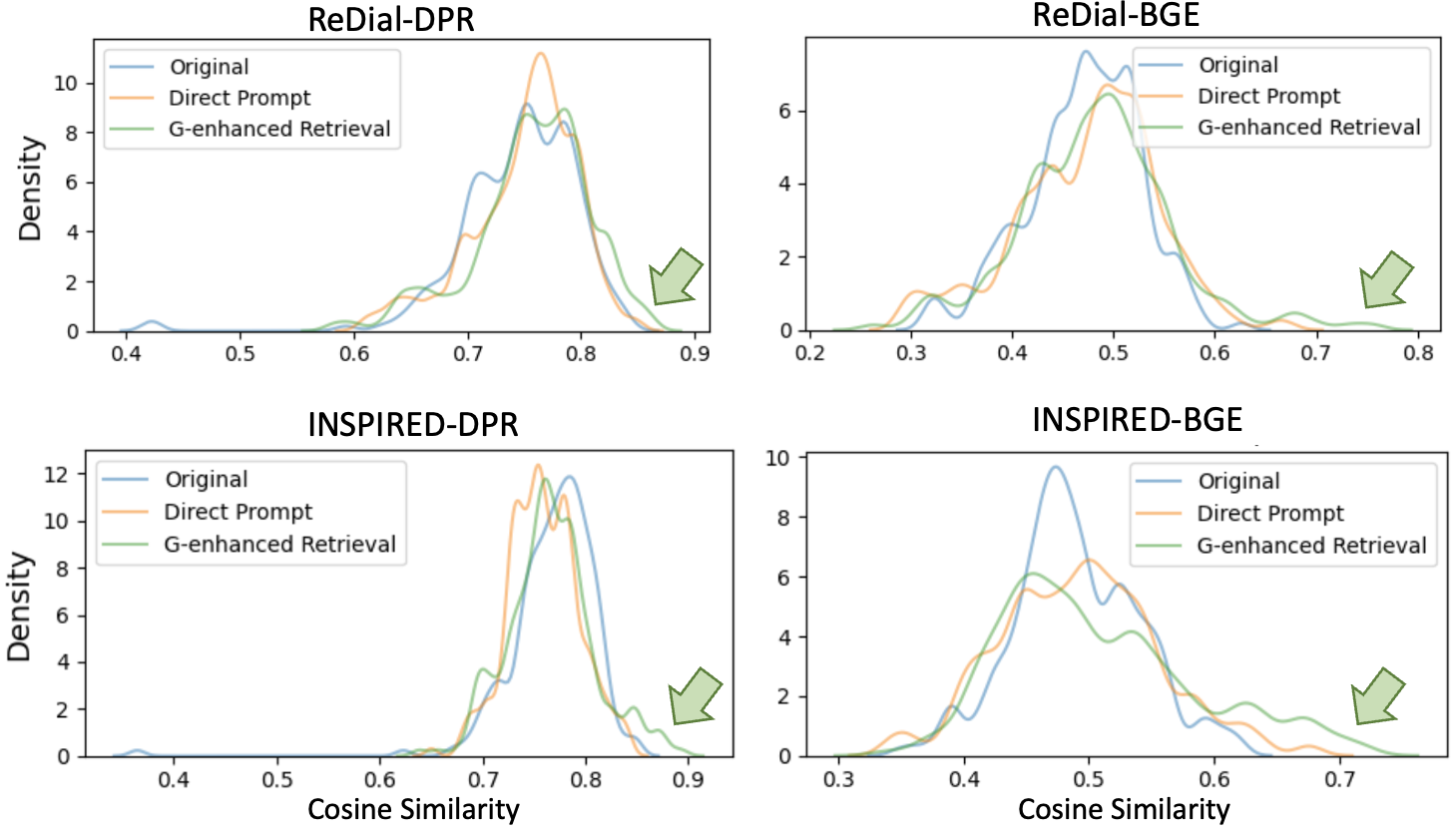}
    \caption{Comparison of query vs. ground-truth cosine similarity distribution (higher is better, Ours is green).}
    \label{fig:add_exp1_1}

\end{figure}

\paragraph{1) Latent Space Alignment.} We analyze the \emph{cosine similarity} distribution between queries and ground-truth items. Because the retrieval module only shows top-ranked items to the generator, queries with higher cosine similarity to the ground-truth item are more likely to yield correct recommendations. We evaluate three query sets: (i) those generated by the $LLM_{QR}$ in G-augmented R, (ii) the original conversation text, and (iii) queries from directly prompting a vanilla LLM (without self-refinement).

Figure~\ref{fig:add_exp1_1} compares the distributions across these three approaches using both DPR and BGE encoders. We observe that G-augmented R yields significantly more queries with higher cosine similarity (note the heavier right tail). This indicates that our self-refinement process produces queries more akin to ground-truth item embeddings—an essential step for retrieving highly relevant candidates.

\paragraph{2) Textual Similarity with Expert-Annotated Queries.} We assessed how closely our queries match linguistic expert annotations by randomly sampling 200 data points from the ReDial and INSPIRED datasets. An expert linguistics annotator received the full conversation context and ground-truth user-preferred item, then wrote a concise query reflecting the user’s needs.

\begin{table}[!ht]
    \centering
    \resizebox{.5\textwidth}{!}{
    \begin{tabular}{c|ccc}
    \hline
        Metric & Raw & Direct Prompt & G-augmented R \\ \hline
        BLEU-1 & 0.0505 & 0.0423 & \textbf{0.0527} \\ 
        BLEU-2 & 0.0155 & 0.0086 & \textbf{0.0184} \\ 
        BLEU-3 & 0.0050 & 0.0016 & \textbf{0.0067} \\ \hline
        ROUGE-1 & 0.1334 & 0.1122 & \textbf{0.1742} \\ 
        ROUGE-2 & 0.0182 & 0.0244 & \textbf{0.0448} \\ 
        ROUGE-L & 0.1157 & 0.0998 & \textbf{0.1529} \\ \hline
    \end{tabular}}
    \caption{Comparison of BLEU and ROUGE scores across queries from different methods.}
    \label{fig:bleu}

\end{table}

We computed BLEU and ROUGE scores to compare our query outputs with these human annotations. As reported in Table~\ref{fig:bleu}, G-augmented R demonstrates higher BLEU and ROUGE scores than both the \textit{raw conversation} text (Raw) and \textit{direct LLM prompting}. The improvement underscores the effectiveness of our generative self-refining approach in capturing the salient details of user intent without discarding critical contextual information—thereby closely mirroring the way humans reformulate queries.

\subsubsection{Retrieval-augmented Generation: Enhance Accuracy on Item Recommendation}

\begin{figure}
    \centering
    \includegraphics[width=1.0\linewidth]{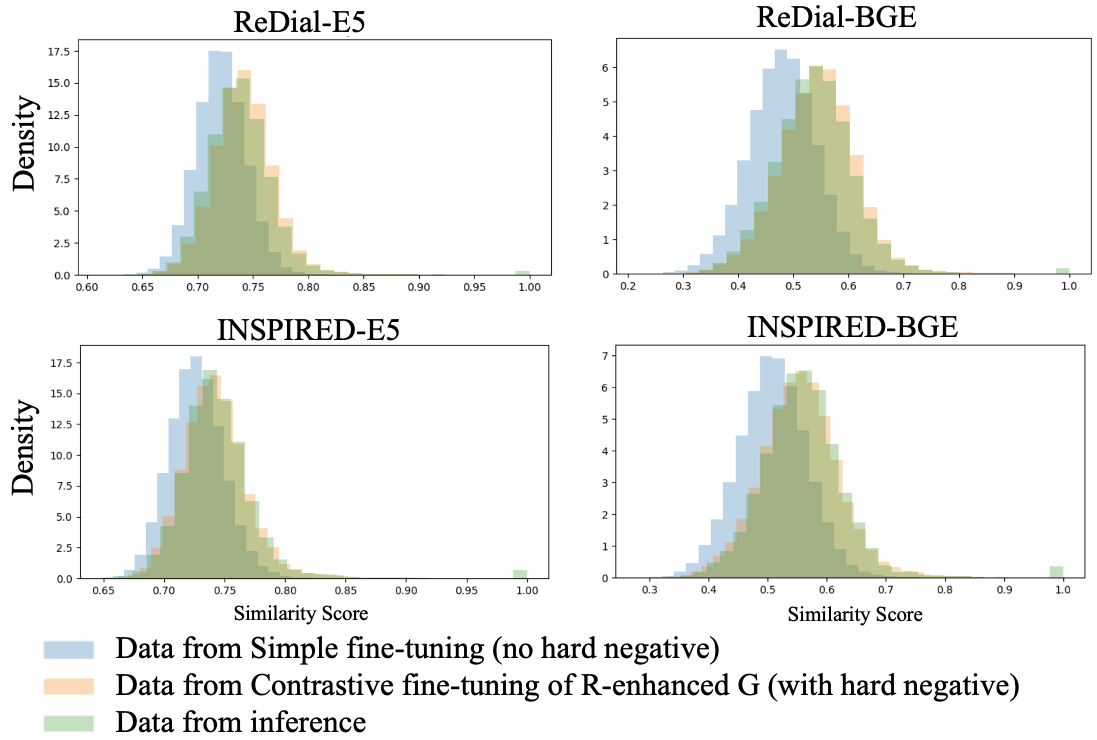}
    \caption{Distribution of LLM's input during training and inference.}
    \label{fig:add_exp2}
\end{figure}

As shown in Figure \ref{fig:add_exp2}, we project the training inputs and inference inputs into a shared embedding space (e.g., using E5~\cite{wang2022text} or BGE~\cite{bge_embedding}). For the random-negative baseline (blue distribution), there is a significant discrepancy between the embeddings of training data and inference data. Conversely, when hard negatives (orange distribution) are used, the alignment between training and inference distributions improves considerably. This alignment explains our observed performance gains: the system trained with hard negatives achieves a higher recommendation success rate because better discrimination between similar items.

Figure \ref{fig:add_exp2} shows how both training and inference inputs are projected into the same embedding space (e.g., using E5~\cite{wang2022text} or BGE~\cite{bge_embedding}). Under the random-negative baseline (blue distribution), there is a pronounced mismatch between the two sets of embeddings. In contrast, using hard negatives (orange distribution) more closely aligns training and inference embeddings. This closer alignment underpins the performance improvements: models trained with hard negatives achieve higher recommendation success rates because they better discriminate among closely related items.

\begin{table}[]
\resizebox{\columnwidth}{!}{
\begin{tabular}{lc|c}
\hline
                           & Model Size          & \multicolumn{1}{l}{Avg Hallu. Ratio} \\ \hline
\multirow{1}{*}{ReGeS (Middle)} & \multirow{1}{*}{8B}            & \textbf{0.13\%}                               \\ \hline
\multirow{2}{*}{LLM-CRS\textsuperscript{\ref{fn:hallu}}}   & $\sim$20B (GPT3.5)        & 4.49\%                               \\
                           & $\sim$200B (GPT4)        & 5.14\%                               \\ \hline
\end{tabular}}
\caption{Comparison of Average Hallucination Ratio.}
\label{table:hallu}

\end{table}

\footnotetext[3]{\label{fn:hallu}Halluciation Ratios of LLM-CRS are directly obtained from~\cite{he2023large}.}

We also compare the hallucination ratio of ReGeS (Middle Computation Setup) with LLM-CRS baselines~\cite{he2023large,hou2024large}. By implementing ReGeS, the hallucination ratio is dramatically decreases from around 5\% in existing LLM-based CRS to under 0.13\%.

Overall, R-augmented G directly addresses the second challenge raised in our introduction—namely, how to refine item generation to handle similar or confusing item candidates. By ensuring the LLM’s training data closely matches its real-world usage scenario, we reduce hallucinations and bolster recommendation accuracy.

\subsection{Discussion}

\subsubsection{Impact of the Number of Retrieved Candidates}
\label{appendix:num_candidates}

A key design choice in retrieval-augmented recommendation is how many retrieved candidates to pass from the retriever to the generation model. Intuitively, presenting fewer items (e.g., 10) can make the generation (i.e., re-ranking) task easier, as there are fewer similars to distinguish. However, fewer candidates risk excluding the correct item more often, especially in real-world scenarios where the retriever is imperfect. Conversely, retrieving more items (e.g., 50) increases coverage but makes the generation step more challenging.

We analyze this trade-off by \emph{forcing} the ground-truth item to appear in the final list. This allows us to isolate the generation model’s behavior via a \textbf{Item Generation Success Rate}: the percentage of times the model actually selects that correct item. 

We also measure the \textbf{Overall Recommendation Success Rate} of the \emph{end-to-end} system, where the retriever may or may not retrieve the correct item. Table~\ref{tab:num_candidates} shows the results for both \textit{ReDial} and \textit{INSPIRED}, comparing two model sizes (2B vs.\ 8B) and two retrievers (DPR vs.\ BGE).

\begin{table}[ht]
\centering
\resizebox{0.98\linewidth}{!}{
\begin{tabular}{l|r|r|r|r}
\hline
\multicolumn{5}{c}{\textbf{ReDial}}\\
\hline
\textbf{Num} & \textbf{Model} & \textbf{Retriever} & \textbf{Overall} & \textbf{Item Generation}\\
\textbf{Candidates} & \textbf{Size} &  & \textbf{Rec Success Rate} & \textbf{Success}\\
\hline
50 & 2B  & DPR     & \textbf{0.1078} & 0.2108 \\
50 & 8B  & BGE     & \textbf{0.1961} & 0.2451 \\
\hline
10 & 2B  & DPR     & 0.0420 & 0.5549 \\
10 & 8B  & BGE     & 0.0593 & 0.5886 \\
\hline
\multicolumn{5}{c}{\textbf{INSPIRED}}\\
\hline
\textbf{Num} & \textbf{Model} & \textbf{Retriever} & \textbf{Overall} & \textbf{Item Generation}\\
\textbf{Candidates} & \textbf{Size} &  & \textbf{Rec Success Rate} & \textbf{Success}\\
\hline
50 & 2B  & DPR     & 0.0543 & 0.1948 \\
50 & 8B  & BGE     & 0.0835 & 0.1894 \\
\hline
10 & 2B  & DPR     & \textbf{0.0784} & 0.3873 \\
10 & 8B  & BGE     & \textbf{0.0846} & 0.4314 \\
\hline
\end{tabular}
}
\caption{
Performance under different final-list sizes. 
\emph{Overall Rec Success Rate} measures the end-to-end system accuracy, including retrieval errors. 
\emph{Item Generation Success Rate} measures how well the generation stage picks the correct item \emph{when it is guaranteed to be in the final list}.
}
\label{tab:num_candidates}
\end{table}

Table~\ref{tab:num_candidates} reveals two interesting dataset-specific insights:
\begin{itemize}
    \item \textbf{ReDial: Retrieval Is the Bottleneck.} 
    ReDial conversations are relatively brief and thus produce less informative queries. Because the retriever’s performance is weaker, using a larger candidate set (\emph{Num Candidates} = 50) substantially increases coverage and leads to higher \emph{Overall Rec Success Rate}. Although selecting the correct item among 50 similar movies is more difficult (yielding a lower Item Generation Success Rate), better coverage outweighs this difficulty in an end-to-end system.
    
    \item \textbf{INSPIRED: Generation Is the Bottleneck.}
    INSPIRED dialogs are longer and contain richer user preferences, producing more precise queries. In this scenario, the retriever performance is strong, so the main challenge becomes fine-grained item distinction during generation. As a result, presenting fewer items (10) actually improves the \emph{Overall Rec Success Rate}, because the generation model faces fewer confusable candidates.
\end{itemize}

In the main experiment, we choose 50 candidates by default, as our Retrieval-Augmented Generation module is specifically designed to distinguish subtle differences between highly similar items. This setting offers broader coverage while maintaining strong item generation accuracy. Nonetheless, practitioners may adjust the candidate-list size based on the expected query quality, desired coverage, and computational constraints.

\subsubsection{Chain-of-Thoughts}

\begin{figure}
    \centering
    \includegraphics[width=1\linewidth]{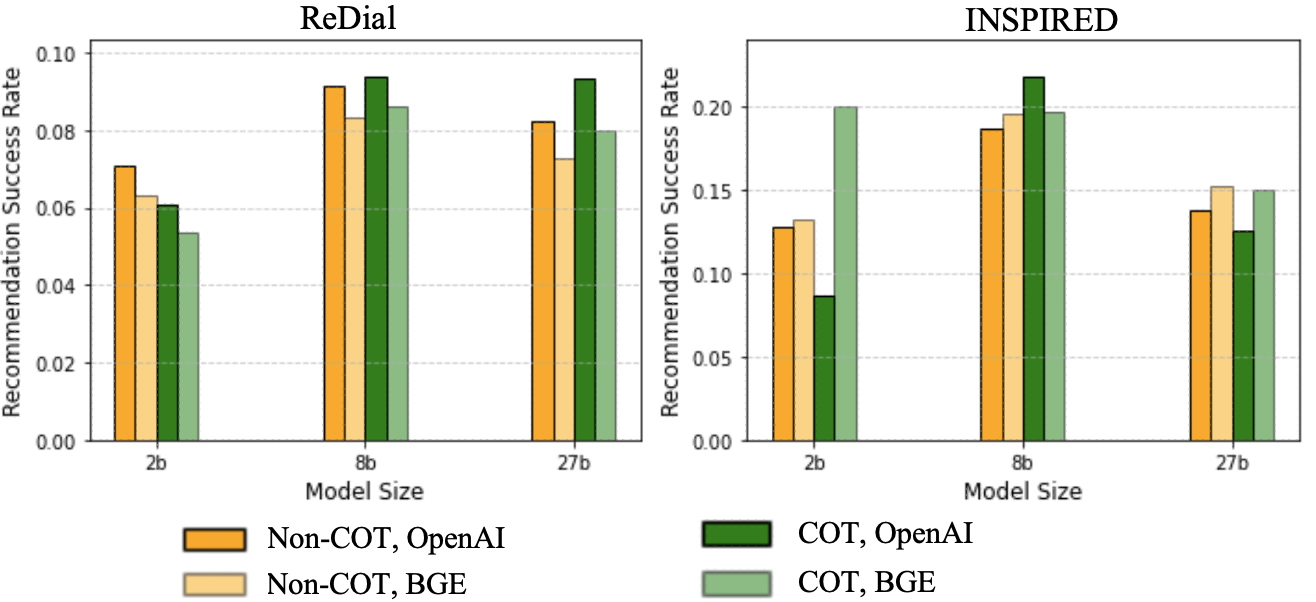}
    \caption{Recommendation performance comparison between Chain-of-Thought and Non-CoT generation.}
    \label{fig:exp3}

\end{figure}

By default, LLM of ReGeS generates the recommended item without providing explanation. However, existing research suggests that chain-of-thought (CoT) generation can enhance the generation quality~\cite{wei2022chain, zhang2022automatic, chu2023survey}. To explore CoT, we fine-tuned LLMs to generate recommendations in a CoT style: first summarizing the user's preferences observed from the user-system interaction history, and then generating the item recommendation.

To create the training data containing reasoning, we prompted the same LLM within the RAG pipeline to summarize user preferences step by step. We then concatenated the CoT reasoning in front of the item recommendation for each training data point. We also add instruction "think step by step". The CoT-style fine-tuned LLM still used hard negatives, just as in the non-CoT R-augmented G.

Figure~\ref{fig:exp3} compares the recommendation success rate between CoT and Non-CoT generations. The results show that CoT-style generation does not universally improve recommendation accuracy. However, when text-embedding-large-3 and Llama3.1-8B are used, CoT-style ReGeS achieves the overall best performance, with a success rate of 0.2179 on INSPIRED and 0.0939 on ReDial, outperforming all other non-CoT generation runs.

\section{Related Work}

Traditional Recommender Systems (RS) primarily rely on implicit signals—such as clicks and ratings—to generate item suggestions~\cite{jannach2021survey}, but often struggle when user interests shift or require nuanced clarifications. By contrast, Conversational Recommender Systems (CRS) employ multi-turn dialogues to iteratively refine user preferences~\cite{zhou2020improving}, yet they must decode subtle, colloquial intents and map them onto structured item data, a task necessitating considerable domain expertise~\cite{gao2021advances}

A popular CRS approach leverages domain-specific models—often graph neural networks over knowledge graphs—to capture user preferences and item attributes~\cite{chen2019towards, zhou2020improving, zhou2021crslab, wang2022towards}. Although effective, these methods frequently rely on rigid graph structures or keyword matching, which can overlook fine-grained linguistic cues. Integrating such representations with generative language models also risks misalignment between recommendation accuracy and conversational fluency. Recently, Large Language Models (LLMs) have been adopted for CRS to exploit their built-in (though static) knowledge and robust language capabilities~\cite{achiam2023gpt, team2024gemma, dubey2024llama, wu2024survey, he2023large}. However, LLMs face two main obstacles: (i) their fixed internal knowledge impedes recommendations for newly introduced items, and (ii) they often hallucinate, suggesting content that does not exist~\cite{lewis2020retrieval, wu2024retrieval, ji2023towards, tonmoy2024comprehensive}.

RAG addresses these issues by combining an external retriever with a generative model to inject up-to-date factual information, thereby reducing reliance on frozen parameters~\cite{shuster2021retrieval, xu2024retrieval, di2023retrieval}. It has shown promise in traditional RS settings~\cite{di2023retrieval, rao2024ramo, ma2024simple}, yet remains underexplored for conversational recommendation.

Unlike prior RAG applications in RS~\cite{di2023retrieval, rao2024ramo, ma2024simple}, ReGeS introduces a reciprocal mechanism specifically addressing the challenges of noisy conversational input (via generation-augmented retrieval) and fine-grained item ambiguity common in CRS (via retrieval-augmented generation). To our knowledge, this is the first RAG-based framework for conversational recommendation, offering a flexible, up-to-date, and hallucination-resistant solution.

\section{Conclusion}

We introduced \textbf{ReGeS}, a novel framework that leverages the reciprocal synergy between retrieval and generation for conversational recommender systems. By integrating Generation-augmented Retrieval with Retrieval-augmented Generation, ReGeS effectively overcomes two key challenges in RAG-based systems: it transforms noisy, multi-turn dialogues into concise, informative queries and refines item selection by contrastively learning from hard negatives. Extensive experiments on the ReDial and INSPIRED datasets show that ReGeS not only nearly doubles the likelihood of delivering a correct recommendation compared to state-of-the-art methods but also dramatically reduces item hallucinations—by over 30 times. Comprehensive ablation studies and analyses further validate the benefits of our reciprocal design from the perspectives of straightforward textual matching and latent distribution alignment. We envision that this work will spur further research into retrieval-augmented strategies for conversational recommendation and inspire future efforts to scale and deploy such systems in real-world applications.

\newpage

\balance
\printbibliography

\end{document}